\documentclass[conference]{IEEEtran}
\IEEEoverridecommandlockouts
\usepackage{cite}
\usepackage{amsmath,amssymb,amsfonts}
\usepackage{algorithmic}
\usepackage{graphicx}
\usepackage[frozencache=true,cachedir=minted]{minted}
\usepackage{textcomp}
\usepackage{xcolor}
\usepackage{hyperref}

\usepackage{pgfplots}
\usepackage{xcolor}
\pgfplotsset{compat=1.16}

\definecolor{mintedrule}{rgb}{0.2,0.2,0.2}
\definecolor{mintedbg}{rgb}{1,1,1}

\setminted{
    frame=lines,
    rulecolor=\color{mintedrule},
    framesep=2mm,
    fontsize=\small,
    baselinestretch=0.92,
    breaklines=true,
    breakanywhere=true,
    autogobble=true,
    linenos=true,
    numbersep=3pt,
    xleftmargin=1.4em,
    tabsize=4,
    escapeinside=||,
    mathescape=true,
    bgcolor=mintedbg,
    highlightcolor=magenta!12
}
\def\BibTeX{{\rm B\kern-.05em{\sc i\kern-.025em b}\kern-.08em
    T\kern-.1667em\lower.7ex\hbox{E}\kern-.125emX}}
\begin{document}

\title{Kotlin-MP: DSL and IR Transformer for Parallelism}

\author{\IEEEauthorblockN{Ruikai Huang}}

\maketitle

\section{Introduction}
While Kotlin provides exceptional abstractions for asynchronous concurrency (like Coroutines for I/O), it lacks low-overhead, directive-based constructs for true hardware parallelism. When developers attempt to parallelize compute-bound mathematical loops using standard task-parallel libraries, they encounter significant runtime overhead such as lambda allocations and state-machine generation due to the abstraction layers, and face difficulties in ensuring readability, reliability and maintainability in implementing them.

To address this, this project proposes Kotlin-MP, a native Kotlin Domain-Specific Language (DSL) paired with a custom Intermediate Representation (IR) transformer as a Kotlin compiler plugin. The DSL provides an intuitive, OpenMP-like syntax for developers to denote parallel regions, schedules, and critical sections. Instead of acting as a standard runtime wrapper, the IR transformer intercepts this DSL at compile-time and structurally rewrites the code. It lowers the parallelized sections directly into optimized \textit{java.util.concurrent.ForkJoinPool} tasks, automatically managing loop chunking, performing thread scheduling, and injecting hardware synchronization and mutual exclusion.

The main contributions of the work are:

\begin{itemize}
    \item A native Kotlin DSL that offers OpenMP-style constructs to achieve parallelism in a readable and idiomatic form.
    \item A Kotlin compiler IR transformer that recognizes the DSL markers and lowers them at compile time into efficient JVM parallel execution patterns.
    \item An evaluation using behavioral tests and JMH benchmarking to show that the implementation produces correct parallel code and scalable performance on compute-bound workloads.
\end{itemize}

\section{Background and Motivation}
There is a fundamental difference between concurrency and parallelism: concurrency is about managing many in-progress tasks; parallelism is about simultaneous execution for speedup. A concurrent program can run on one core via interleaving, while a parallel program needs hardware resources that can execute multiple operations at once~\cite{distinction}.
Kotlin's flagship feature, \textit{Coroutines}~\cite{kcoroutines}, relies on \textit{Continuation-Passing Style} (CPS) state machines. While exceptional for I/O concurrency, CPS introduces unacceptable instruction overhead when applied to tight, CPU-bound loops.
Figure~\ref{fig:sequential-matmul} presents a typical sequential implementation of matrix multiplcation in Kotlin: it uses three nested loops to perform the computation. We can apply coroutines to accelarate computation. As shown in Figure~\ref{fig:corountine-matmul}, the outermost loop \textit{i} is processed concurrently (and likely parallized, depending on the runtime environment and hardware constraint) by assigning every \textit{i} a coroutine. While the abstraction is straightforward and expressive, the overhead on the lambda construction, scheudling, and state-management is not ideal for dense compute-bound kernels.

\begin{figure}[t]
\centering
\begin{minted}[baselinestretch=0.65,fontsize=\scriptsize]{java}
for (i in 0 until size) {
  for (j in 0 until size) {
    var sum = 0.0f
    for (k in 0 until size) {
      sum += matrixA[i * size + k] * matrixB[k * size + j]
    }
    matrixC[i * size + j] = sum
  }
}
\end{minted}
\caption{Sequential matrix multiplication implementation.}
\label{fig:sequential-matmul}
\vspace{-1.2em}
\end{figure}

\begin{figure}[t]
\centering
\begin{minted}[baselinestretch=0.65,fontsize=\scriptsize]{java}
runBlocking(Dispatchers.Default) {
  (0 until size).map { i ->
    async {
      for (j in 0 until size) {
        var sum = 0.0f
        for (k in 0 until size) {
          sum += matrixA[i * size + k] * matrixB[k * size + j]
        }
        matrixC[i * size + j] = sum
      }
    }
  }.awaitAll()
}
\end{minted}
\caption{Kotlin coroutine-based matrix multiplication implementation.}
\label{fig:corountine-matmul}
\vspace{-2.2em}
\end{figure}

To achieve true hardware parallelism on the JVM, the optimal target is Java's \textit{ForkJoinPool}~\cite{forkjoin}, which utilizes a specialized Cilk-style~\cite{cilk} work-stealing algorithm designed specifically to keep physical cores saturated during data-parallel operations. However, writing raw \textit{ForkJoinPool} boilerplate is verbose and error-prone. Figure~\ref{fig:forkjoin-matmul} illustrates a \textit{ForkJoin}-based implementation for matrix multiplcation code with the first layer parallelized. The developer has to manually manage the thread pool, construct subtasks, associate substasks to threads, and perform task submission and synchronization. This method translates directly to JVM-level multithreading and thereby reduces runtime overhead, but comes with the heavy erro-prone boilerplate.

\begin{figure}[t]
\centering
\begin{minted}[baselinestretch=0.65,fontsize=\scriptsize]{java}
val pool = ForkJoinPool.commonPool()
val numThreads = pool.parallelism.coerceAtLeast(1)
val cSize = (size + numThreads - 1) / numThreads
val tasks =
  (0 until numThreads).mapNotNull { threadId ->
    val chunkStart = threadId * cSize
    if (chunkStart >= size) return@mapNotNull null
    val chunkEnd = minOf(chunkStart + cSize - 1, size-1)
    Runnable {
      for (i in chunkStart..chunkEnd) {
        for (j in 0 until size) {
          var sum = 0.0f
          for (k in 0 until size) {
            sum += matrixA[i * size + k] * matrixB[k * size + j]
          }
          matrixC[i * size + j] = sum
        }
      }
    }
  }
val futures = tasks.map { pool.submit(it) }
futures.forEach { it.get() }
\end{minted}
\caption{ForkJoin-based matrix multiplication implementation.}
\label{fig:forkjoin-matmul}
\end{figure}

By applying the compiler transformation to the idiomatic OpenMP-style Kotlin DSL, this project bridges the gap. It provides developers with zero-cost syntactic abstractions while forcing the compiler to generate the optimized, native thread-mapping bytecode required for true parallelism. Figure~\ref{fig:kotlinmp-matmul} demonstrates how the same thing can be achieved using Kotlin-MP. The DSL preserves the readability at the high-level by avoiding boilerplates, while the compiler plugin lowers the program into an efficient parallel execution.

\begin{figure}[t]
\centering
\begin{minted}[baselinestretch=0.65,fontsize=\scriptsize]{java}
omp {
  parallelFor(0 until size) { i ->
    for (j in 0 until size) {
      var sum = 0.0f
      for (k in 0 until size) {
        sum += matrixA[i * size + k] * matrixB[k * size + j]
      }
      matrixC[i * size + j] = sum
    }
  }
}
\end{minted}
\caption{Kotlin-MP matrix multiplication implementation.}
\label{fig:kotlinmp-matmul}
\vspace{-1.2em}
\end{figure}


\begin{figure}[t]
\centering
\includegraphics[width=0.9\columnwidth,page=1]{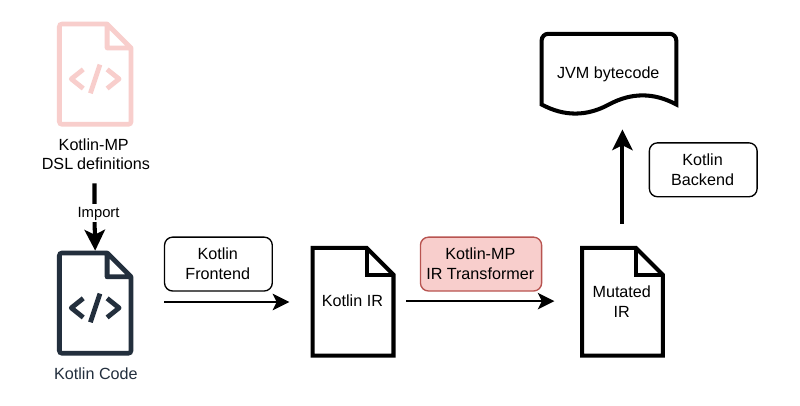}
\caption{Overview of the Kotlin-MP Approach}
\label{fig:kotlinmp-workflow}
\vspace{-1.2em}
\end{figure}

\section{Kotlin-MP Approach}
The Kotlin-MP approach has two components: (1) a DSL implemented using \textit{Type-Safe Builder}, acting as syntactic markers, and (2) a Kotlin Compiler Plugin (i.e., an IR transformer) to perform the lowering. Figure~\ref{fig:kotlinmp-workflow} presents the overview of this approach. To use Kotlin-MP, the developer first needs to import the DSL definitions as a library. In the code, the developer leverages the DSL to construct parallelism blocks. Then, the developer compile the source code using the Kotlin compiler with the Kotlin-MP plugin attached. During the compilation, once the compiler frontend parsed the source code into Kotlin IR, the compiler plugin transforms the IR, rewriting DSL constructs with plain \textit{ForkJoin} multithreading implementation. Eventually, the compiler backend generates executable JVM bytecode files from the mutated IR. The Kotlin-MP implementation is \href{https://github.com/nntzuekai/Kotlin-MP}{open-sourced}.

\subsection{DSL Markers}

The DSL serves as a Kotlin library that can be imported into developer projects. It provides the following constructs and clauses in the OpenMP style.

\textit{omp\{...\}}. Denotes a code block within which the developer can use Kotlin-MP to define parallelism code.

\textit{parallelFor\{...\}}. A parallelized for loop, with the range and the index variable of the loop specified. Optionally, the developer can specify the schedule (i.e., static or dynamic) and the number of threads they want to use.

\textit{parallel\{...\}}. Similar to \textit{parallelFor\{...\}}, but a general code block to be parallezied.

\textit{critical\{...\}} and \textit{critical(name)\{...\}}. A critical code section in the parallelized code block, with an optional name, used for mutual exclusion.

\textit{barrier()}. A barrier in the parallelized code block, used for global synchronization. Only allowed in \textit{parallel\{...\}}.

The DSL itself provides a dummy implementation for these constructs. In the case the compiler plugin is absent or not invoked, the dummy implementation ensures the correctness of the program by executing sequentially.

\subsection{IR Transformer}

The IR transformer works at IR-level to rewrite all occurences of the DSL markers with their corresponding \textit{ForkJoin} implementation. Although the DSL markers syntatically introduce lambda allocation and invocation overhead at the source code level, this IR transformer optimizes and inlines the actual implementation as much as possible, eliminating unnecessary runtime overhead in the generated bytecode.

\textbf{\textit{parallelFor}}. They will be lowered to an efficient \textit{ForkJoin} implementation, as in Figure~\ref{fig:forkjoin-matmul}. Based on the schedule and the number of threads, subtasks will be created and associated with threads, and then submitted to the thread pool for execution. The scheduling strategy is in OpenMP style: for static schedule, when no chunk size is specified, it will calculate the chunk size such that each available thread gets just one chunk; when the chunk size is specified for static schedule, threads are allocated to chunks in a pre-determined order; for dynamic schedule, a work-queue mechanism is used: the first available thread gets the next chunk to process. By default, \textit{parallelFor} uses the strategy of static schedule with no chunk size.

\textbf{\textit{parallel}}. Similar to \textbf{\textit{parallelFor}}, but schedule is not used. The \textit{parallel} block will be executed in parallel with the specified number of threads. When the thread number is not specified, it used the maximum number of available threads.



\textbf{\textit{critical}}. The IR transformer will inject usage of a mutex lock to guard the enclosed code block, ensuring that at most one thread can enter the code block at any time. The critical block can be optionally named for fine-grained control of mutual exclusion. The name of critical blocks is effective globally: even in different blocks, critical blocks with the same name (or unnamed) are guarded by the same mutex lock.


\textbf{\textit{barrier}}. This is only allowed in \textit{parallel} as \textit{parallelFor} already comes with a global synchronization at the end of the loop. When the IR transformer detects that \textit{barrier} is used within a \textit{parallel}, it will place a barrier object at the corresponding place to enforce synchronization.


\section{Evaluation}

\subsection{Behavioral Unti Tests}
We created a test suite to verify the correctness of all implemeneted features. Kotlin-MP implementation passed all unit tests, proving that the IR transformers produce correct parallelism code at compile-time. Figure~\ref{fig:critical-test} shows an example of unit tests: it verifies the effectiveness of \textit{critical} by checking against an intentional data race.

\begin{figure}[t]
\centering
\begin{minted}[baselinestretch=0.65,fontsize=\scriptsize]{java}
omp {
  parallelFor(0 until size, Schedule.Dynamic(100)) {
    //Data race (lose counts due to thread collisions)
    unprotectedCounter++
    //Unnamed critical (perfectly safe)
    critical { protectedCounter++ }
    //Named critical (perfectly safe, independent lock)
    critical("CounterLock") {
      namedProtectedCounter++
    }
  }
}
\end{minted}
\caption{Partial digest of the unit test to verify the implementation of \textit{critical}}
\label{fig:critical-test}
\vspace{-1.2em}
\end{figure}

\subsection{Performance and Scaling Studies}

We applied \textit{Java Microbenchmark Harness} (JMH) to evaluate the performance and scalability of Kotlin-MP on different computational tasks. Due to the page limit, we here present the results of two representative benchmarks: (1) \textbf{Float Matrix Multiplcation} for comparison of runtime overhead across different methods to implement parallelism in Kotlin, and (2) \textbf{Irregular Load Task}, which evaluates the correctness and runtime overhead of dynamic scheduling.

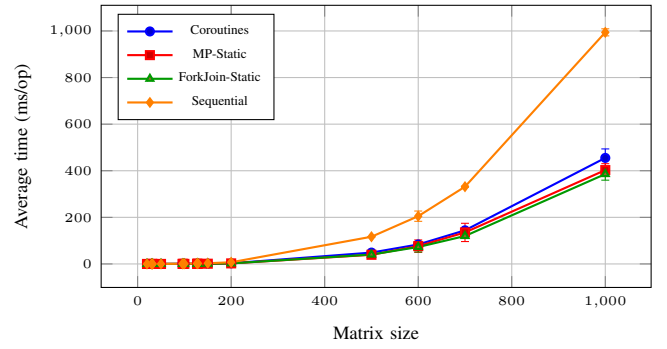
\begin{figure}[t]
\centering
\begin{tikzpicture}
\begin{axis}[
    width=\columnwidth,
    height=0.6\columnwidth,
    xlabel={Matrix size},
    ylabel={Average time (ms/op)},
    grid=both,
    legend pos=north west,
    legend style={font=\tiny},
    tick label style={font=\tiny},
    label style={font=\scriptsize},
    ytick={0,200,400,600,800,1000},
]

\addplot+[
    color=blue,
    mark=*,
    mark size=1.5pt,
    thick,
    error bars/.cd,
    y dir=both,
    y explicit
] coordinates {
    (20,0.027) +- (0,0.001)
    (30,0.041) +- (0,0.003)
    (31,0.043) +- (0,0.004)
    (32,0.044) +- (0,0.001)
    (50,0.082) +- (0,0.001)
    (96,0.453) +- (0,0.132)
    (97,0.426) +- (0,0.036)
    (100,0.484) +- (0,0.114)
    (127,0.915) +- (0,0.067)
    (128,0.871) +- (0,0.050)
    (129,0.835) +- (0,0.092)
    (150,1.256) +- (0,0.044)
    (200,2.810) +- (0,0.056)
    (500,48.692) +- (0,0.974)
    (600,83.522) +- (0,1.450)
    (700,144.260) +- (0,2.347)
    (1000,455.101) +- (0,38.608)
};
\addlegendentry{Coroutines}

\addplot+[
    color=red,
    mark=square*,
    mark size=1.5pt,
    thick,
    error bars/.cd,
    y dir=both,
    y explicit
] coordinates {
    (20,0.015) +- (0,0.005)
    (30,0.026) +- (0,0.006)
    (31,0.028) +- (0,0.002)
    (32,0.026) +- (0,0.001)
    (50,0.056) +- (0,0.001)
    (96,0.275) +- (0,0.006)
    (97,0.294) +- (0,0.012)
    (100,0.314) +- (0,0.003)
    (127,0.668) +- (0,0.262)
    (128,0.728) +- (0,0.010)
    (129,0.685) +- (0,0.111)
    (150,1.008) +- (0,0.009)
    (200,2.354) +- (0,0.063)
    (500,39.289) +- (0,1.071)
    (600,76.471) +- (0,26.560)
    (700,135.277) +- (0,39.186)
    (1000,402.776) +- (0,27.591)
};
\addlegendentry{MP-Static}

\addplot+[
    color=green!60!black,
    mark=triangle*,
    mark size=1.5pt,
    thick,
    error bars/.cd,
    y dir=both,
    y explicit
] coordinates {
    (20,0.012) +- (0,0.001)
    (30,0.023) +- (0,0.004)
    (31,0.025) +- (0,0.001)
    (32,0.026) +- (0,0.001)
    (50,0.057) +- (0,0.004)
    (96,0.285) +- (0,0.040)
    (97,0.285) +- (0,0.010)
    (100,0.321) +- (0,0.004)
    (127,0.625) +- (0,0.007)
    (128,0.731) +- (0,0.020)
    (129,0.646) +- (0,0.008)
    (150,1.005) +- (0,0.011)
    (200,2.387) +- (0,0.363)
    (500,40.322) +- (0,5.435)
    (600,72.617) +- (0,21.280)
    (700,119.822) +- (0,2.364)
    (1000,385.827) +- (0,26.275)
};
\addlegendentry{ForkJoin-Static}

\addplot+[
    color=orange,
    mark=diamond*,
    mark size=1.5pt,
    thick,
    error bars/.cd,
    y dir=both,
    y explicit
] coordinates {
    (20,0.007) +- (0,0.001)
    (30,0.022) +- (0,0.001)
    (31,0.024) +- (0,0.001)
    (32,0.027) +- (0,0.002)
    (50,0.105) +- (0,0.011)
    (96,0.794) +- (0,0.116)
    (97,0.796) +- (0,0.007)
    (100,0.893) +- (0,0.009)
    (127,1.807) +- (0,0.017)
    (128,1.994) +- (0,0.029)
    (129,1.918) +- (0,0.012)
    (150,3.058) +- (0,0.033)
    (200,7.136) +- (0,0.154)
    (500,116.735) +- (0,0.671)
    (600,205.029) +- (0,22.083)
    (700,331.547) +- (0,2.673)
    (1000,994.084) +- (0,15.118)
};
\addlegendentry{Sequential}

\end{axis}
\end{tikzpicture}
\caption{Comparing the execution time of float matrix multiplication across different parallelism methods as the matrix size increases.}
\label{fig:float-linear-tikz}
\vspace{-1.2em}
\end{figure}

\textbf{Float Matrix Multiplcation}: This benchmark compares the execution time of matrix multiplcation of float type across different implementations by matrix sizes. Results are shown in Figure~\ref{fig:float-linear-tikz}.
\textit{Sequential} uses the plain three-loop algorithm. 
The other three parallelized implementation applied different methods to parallelize the first layer loop: 
\textit{Coroutines} uses Kotlin Coroutine; 
\textit{ForkJoin-Static} is the manually written \textit{ForkJoin}-based multithreading; 
\textit{MP-Static} leverages Kotlin-MP's \textit{parallelFor} with default settings (static scheduling, automatic chunk size). 
As shown in Figure~\ref{fig:float-linear-tikz}, all three parallelized versions significantly outperformed \textit{Sequential}, among which \textit{Coroutines} in general had the largest execution time due to its runtime overhead. \textit{MP-Static} achieved performance on a par with \textit{ForkJoin-Static}, and the minor difference is due to that a lambda is not fully inlined by the compiler backend.


\textbf{Irregular Load Task}: This benchmark is designed to evaluate the correctness and efficiency of Kotlin-MP's dynamic scheduling. It spawns a number of subtasks with highly imbalanced workload on them. With static scheduling, threads with heavier workload will significantly slow down the overall execution time. 
As shown in Table~\ref{irregular_comp}, we evaluated implementations with Kotlin Coroutine, Kotlin-MP dynamic scheduling without and with chunking (\textit{MP-Dyn} and \textit{MP-Dyn-Chunk}), Kotlin-MP static scheduling (\textit{MP-Static}), manually written \textit{ForkJoin} with dynamic and static scheduling, and sequential loops. 
We focus on the comparison between \textit{MP-Dyn} and \textit{ForkJoin-Dyn}: \textit{MP-Dyn} actually achieved slightly better performance overall, with a minor disadvantage only on 1,000 subtasks. Notably, \textit{MP-Dyn-Chunk} shows best performance on larger problems with the help of chunking.

\begin{table}[!ht]
    \centering
    \caption{Execution time of the irregular load benchmark across parallelism implementations on different problem sizes.}
    \begin{tabular}{l|llllll}
    \hline
        Implementation & 1,000 & 3,000 & 5,000 & 10,000 & 15,000 & 30,000 \\ \hline
        Coroutines & 0.583 & 1.674 & 2.769 & 5.603 & 8.400 & 17.147 \\ 
        \textbf{MP-Dyn} & \textbf{0.137} & \textbf{0.388} & \textbf{0.634} & \textbf{1.248} & \textbf{1.870} & \textbf{\phantom{0}3.740} \\ 
        MP-Dyn-Chunk & 0.138 & 0.342 & 0.566 & 1.097 & 1.638 & \phantom{0}3.287 \\ 
        MP-Static & 0.582 & 1.681 & 2.774 & 5.523 & 8.250 & 16.562 \\ 
        \textbf{ForkJoin-Dyn} & \textbf{0.119} & \textbf{0.374} & \textbf{0.640} & \textbf{1.257} & \textbf{1.896} & \textbf{\phantom{0}3.725} \\ 
        ForkJoin-Static & 0.581 & 1.676 & 2.784 & 5.517 & 8.266 & 16.501 \\ 
        Sequential & 0.564 & 1.688 & 2.818 & 5.637 & 8.434 & 16.874 \\ \hline
    \end{tabular}
    \label{irregular_comp}
\vspace{-1.2em}
\end{table}





\section{Related Work}

OpenMP~\cite{OpenMP} defines the canonical scheduling, barrier, and critical section semantics that we aim to support in Kotlin. Java's Fork/Join framework~\cite{forkjoin} details the work-stealing runtime model that the IR transformations target, while Cilk~\cite{cilk} provides the broader historical foundation for lightweight task parallelism and divide-and-conquer scheduling.

On the JVM, JOMP~\cite{JOMP}, JCilk~\cite{JCilk}, Habanero-Java~\cite{HJ}, and X10~\cite{X10} show that directive-style, task-parallel, and richer language-level abstractions can all be adapted to managed runtimes.

These systems collectively show that high-level parallel abstractions can improve programmability, but they generally introduce either new language models or specialized runtime systems. 
In contrast, Kotlin-MP focuses on an OpenMP-style embedded DSL and compiler-plugin-based lowering targeting plain JVM execution through \textit{ForkJoinPool}. 
Prior work on Kotlin coroutines~\cite{kcoroutines} provides the baseline for comparison, as it analyzes the CPS whose overhead this work seeks to avoid in compute-bound parallel loops. 
Complementing that implementation-focused view, Zieliński~\cite{Zielinski2020} compares Kotlin coroutines with Java and Scala solutions for parallel programming, further motivating a careful evaluation of Kotlin-specific parallel abstractions.

\bibliographystyle{IEEEtran}
\bibliography{references}
\end{document}